%   Compilare in PlainTeX

%%%%%%%%%%%%%%%%%%%%%%%%  Format %%%%%%%%%%%%%%%%%%%%%%%%%%%%%%
\newcount\notenumber

\def\note{\global\advance\notenumber by 1
\footnote{{\mathsurround=0pt$^{\the\notenumber}$}}
}

\mag=1300
\normalbaselineskip=11.66pt
\normallineskip=2pt minus 1pt
\normallineskiplimit=1pt
\normalbaselines
\vsize=17.3cm
\hsize=12.1cm
\parindent=20pt
\smallskipamount=3.6pt plus 1pt minus 9pt
\abovedisplayskip=1\normalbaselineskip plus 3pt minus 9pt
\belowdisplayskip=1\normalbaselineskip plus 3pt minus 9pt
\skip\footins=2\baselineskip
\advance\skip\footins by 3pt

\mathsurround 2pt
\newdimen\leftind \leftind=0cm
\newdimen\rightind \rightind=0.65cm

\def\pagenumbers{\footline={\hss\tenrm\folio\hss}}
\nopagenumbers
%\def\folio{\ifnum\pageno<0\uppercase\expandafter
%{\romannumeral-\pageno}\else\number\pageno\fi}

%\headline={\ifodd\pageno\rightheadline\else\leftheadline\fi}

%\def\rightheadline{\phantom{\ninerm\folio}\hfil{\sevenrm\testatinatitolo}\hfil\ninerm\folio}
%\def\leftheadline{\ninerm\folio\hfil{\sevenrm\testatinautore}\hfil\phantom{\ninerm\folio}}

%\def\testatinatitolo{}
%\def\testatinautore{}

\font\sevenrm=cmr7
\font\seventeenbf=cmbx12 at 14pt
%\font\seventeenbf=TimesB at 17pt

\def\MainHead{{\baselineskip 16.7pt\seventeenbf
\noindent\titolo\par}
\normalbaselines
\vskip 28.34pt
\noindent\autori\par
\ni\indirizzo
\footnote{\phantom{i}}{\piedipagina}
\vskip 93.34pt
\ni {\bf Abstract.} \Abstract
\np
}

\font\tenrm=cmr10
\font\tenit=cmti10
\font\tensl=cmsl10
\font\tenbf=cmbx10
\font\tentt=cmtt10

\def\tenpoint{%
\def\rm{\fam0\tenrm}%
\def\it{\fam\itfam\tenit}%
\def\sl{\fam\slfam\tensl}%
\def\tt{\fam\ttfam\tentt}%
\def\bf{\fam\bffam\tenbf}%
}

\tenpoint\rm

%%%%%%%%%%%%%%%%%%%%%%%%  AC.tex %%%%%%%%%%%%%%%%%%%%%%%%%%%%%%
%
%   These Macros provides automatic counters in PlainTeX for: 
%	      Formulae
%       Chapters
%       Sections
%       Subsections
%       Theorems, Propositions, ect.
%       
%
%
%   Fonts definitions
%
\font\ChapTitle=cmbx12% at 17pt
\font\SecTitle=cmbx12% at 12pt
\font\SubSecTitle=cmbx12% at 12pt
%
%
%   Indent definitions
%
\def\NormalSkip{\parskip=5pt\parindent=15pt\baselineskip=12pt\PageNumbers%
\leftskip=0cm\rightskip=0cm}

\def\PageNumbers{\footline={\hss\tenrm\folio\hss}}
\def\NoPageNumbers{\footline={}}

%
%
%   Skips definitions
%
\def\np{\vfill\eject}
\def\ss{\vskip 5pt}
\def\ms{\vskip 15pt}
\def\bs{\vskip 30pt}

\def\ni{\noindent}
%
%
%   Counters definitions
%
\newcount\CHAPTER		          %
\newcount\SECTION		          %
\newcount\SUBSECTION		       %
\newcount\FNUMBER     % Define \FNUMBER to be the counter for formulae
%
%
%   Counters initializations
%
\CHAPTER=0		          % Initialize \CHAPTER to 0
\SECTION=0		          % Initialize \SECTION to 0
\SUBSECTION=0		       % Initialize \SUBSECTION to 0
\FNUMBER=0		          % Initialize \FNUMBER to 0
%
%
% Definition of Plain style for chapters numbering
%
\long\def\NewChapter#1{\global\advance\CHAPTER by 1%
\np\NoPageNumbers\ \vfil\parindent=0cm\leftskip=2cm\rightskip=2cm{\ChapTitle #1\hfil}%
\vskip 4cm\ \vfil\eject\SECTION=0\SUBSECTION=0\FNUMBER=0\NormalSkip}
%
% 
% Definition of Plain style for sections numbering
%
\gdef\Mock{}
\long\def\NewSection#1{\bs\global\advance\SECTION by 1%
\ni{\SecTitle \ifnum\CHAPTER>0 \the\CHAPTER.\fi\the\SECTION.\ #1}\ms\SUBSECTION=0\FNUMBER=0}
\def\CurrentSection{\global\edef\Mock{\the\SECTION}} %
%
% Definition of Plain style for subsections numbering
%
\long\def\NewSubSection#1{\global\advance\SUBSECTION by 1%
{\SubSecTitle \ifnum\CHAPTER>0 \the\CHAPTER.\fi\the\SECTION.\the\SUBSECTION.\ #1}\ss\FNUMBER=0}
%
%
% Definition of Plain style for formulae numbering
%
%
\def\RightFormulaNumber{\global\advance\FNUMBER by 1\eqno{\fopen\the\FNUMBER\fclose}}
\def\LeftFormulaNumber{\global\advance\FNUMBER by 1\leqno{\fopen\the\FNUMBER\fclose}}
\def\RightFormulaLabel#1{\global\advance\FNUMBER by 1%
\eqno{\fopen\the\FNUMBER\fclose}\global\edef#1{\fopen\the\FNUMBER\fclose}}
\def\LeftFormulaLabel#1{\global\advance\FNUMBER by 1%
\leqno{\fopen\the\FNUMBER\fclose}\global\edef#1{\fopen\the\FNUMBER\fclose}}
%
%
% Definition of Composed style for formulae numbering
%
\def\HeadNumber{\ifnum\CHAPTER>0 \the\CHAPTER.\fi%
\ifnum\SECTION>0 \the\SECTION.\ifnum\SUBSECTION>0 \the\SUBSECTION.\fi\fi}
\def\Getfn{\global\advance\FNUMBER by 1 {\fopen\HeadNumber\the\FNUMBER\fclose}}
\def\Getfl#1{\global\advance\FNUMBER by 1 {\fopen\HeadNumber\the\FNUMBER\fclose}%
\global\edef#1{\fopen\HeadNumber\the\FNUMBER\fclose}}
\def\ComposedRightFormulaNumber{\global\advance\FNUMBER by 1%
\eqno{\fopen\HeadNumber\the\FNUMBER\fclose}}
\def\ComposedLeftFormulaNumber{\global\advance\FNUMBER by 1%
\leqno{\fopen\HeadNumber\the\FNUMBER\fclose}}
\def\ComposedRightFormulaLabel#1{\global\advance\FNUMBER by 1%
\eqno{\fopen\HeadNumber\the\FNUMBER\fclose}%
\global\edef#1{\fopen\HeadNumber\the\FNUMBER\fclose}}
\def\ComposedLeftFormulaLabel#1{\global\advance\FNUMBER by 1%
\leqno{\fopen\HeadNumber\the\FNUMBER\fclose}%
\global\edef#1{\fopen\the\FNUMBER\fclose}}
%
%
%  Definition of Plain style for Theorem Numbering 
%
\def\TheoremNumber{\global\advance\FNUMBER by 1 \fopen\the\FNUMBER\fclose}
\def\TheoremLabel#1{\global\advance\FNUMBER by 1\fopen\the\FNUMBER\fclose%
\global\edef#1{\fopen\the\FNUMBER\fclose}}
%
%
%  Definition of Composed style for Theorem Numbering 
%
\def\ComposedTheoremNumber{\global\advance\FNUMBER by 1 \fopen\HeadNumber\the\FNUMBER\fclose}
\def\ComposedTheoremLabel#1{\global\advance\FNUMBER by 1\fopen\HeadNumber\the\FNUMBER\fclose%
\global\edef#1{\fopen\HeadNumber\the\FNUMBER\fclose}}
%
%
%     Style definition
%
\def\fopen{(}\def\fclose{)}                 % Define Formula Number Delimiters
\def\fn{\ComposedRightFormulaNumber}        % Define Formula Number Style
\def\fl{\ComposedRightFormulaLabel}         % Define Formula Label  Style
             % Define Theorem Number Style
              % Define Theorem Label  Style

\def\Compare#1#2{\message{^^J Compara \noexpand #1:=#1[#2]^^J}}

%%%%%%%%%%%%%%%%%%%%%%%%%%  BIBLIO.tex  %%%%%%%%%%%%%%%%%%%%%%%%%%%%%
%%%%%%%%
%
%  -It provides macros for references
%   A file with \bib{}{} declarations should be \input at the beginning
%   Then one can refer to them by \ref{}
%   References are stored in \refs{} (in order of quotation) and displayd by \Biblio
%
%       \bib{tokenname}{Title}
%       \ref{tokenname}
%       \refs
%       \Biblio
%       \LogRef{tokenname}
%
%       \eightpoint
%       \tenpoint
%
%%%%%%%%

\def\ni{\noindent}
\def\ss{\vskip 5pt}
\def\ms{\vskip 10pt}

\def\noex{\noexpand}

\def\refs{}
\def\empty{\#}
\def\BibNumber{}
\def\BibTitle{}

\newcount\BNUM
\BNUM=0

\def\bib#1#2{\gdef#1{\global\def\BibNumber{\empty}\global\def\BibTitle{#2}}}

\def\ref#1{#1%%%%
\if\BibNumber\empty \global\advance\BNUM 1%%%%
\message{reference[\BibNumber]}\message{}%%%
\global\edef\refs{\refs \ss\ni[\the\BNUM]\ \BibTitle}%%%%
\global\edef#1{\noex\global\noex\edef\noex\BibNumber{[\the\BNUM]}%%%%
 \noex\global\noex\edef\noex\BibTitle{\BibTitle}}%%%%
{\bf [\the\BNUM]}%%%%
\else%%%%
{\bf \BibNumber}%%%%
\fi}

\def\Biblio{{\refs}}

%%%%%%%%%%%%%%%%%%%%%%%%%%%%  MACROS %%%%%%%%%%%%%%%%%%%%%%%%%%%%%

\def\Lor{\hbox{\rm Lor}}

\def\Div{\hbox{\rm Div}}

\def\ds{\hbox{\bf ds}}
\def\d{\hbox{\rm d}}
\def\dt{\hbox{\rm dt}}
\def\dr{\hbox{\rm dr}}

\def\dim{\hbox{\rm dim}}

\def\Lie{\hbox{\it \$}}

\def\calE{{\cal E}} 
 
\def\calL{{\cal L}} 
\def\calW{{\cal W}} 
\def\calJ{{\cal J}} 
\def\calU{{\cal U}} 
\def\calB{{\cal B}} 
 
\def\calS{{\cal S}} 
 
\def\calN{{\cal N}}

%            Greek

\def\La{\Lambda}
\def\na{\nabla}
\def\la{\lambda}
\def\Si{\Sigma}
\def\si{\sigma}
\def\ka{\kappa}
\def\Om{\Omega}
\def\om{\omega}
\def\ep{\epsilon}
\def\al{\alpha}
\def\be{\beta}
\def\Ga{\Gamma}
\def\ga{\gamma}

\def\de{\delta}
\def\ze{\zeta}
\def\te{\theta}

\def\vp{\varphi}

\def\R{{\Bbb R}}

%            Mat

\def\R{I \kern-.36em R}
\def\E{I \kern-.36em E}
\def\F{I \kern-.36em F}
\def\Co{I \kern-.66em C}
\def\id{1 \kern-.36em I}              %  Identita'

\def\del{\partial}                   %  Derivata parziale
                      %  Infinito
                   %  Intersezione
                  %  Unione disgiunta
                   %  Unione
            %  Valore assoluto

\def\QDE{{\offinterlineskip\lower1pt\hbox{\kern2pt\vrule width0.8pt%%%%
\vbox to8pt{\hbox to6pt{\leaders\hrule height0.8pt\hfill}\vfill%
\hbox to6pt{\hrulefill}}\vrule\kern3pt}}}
            %  aggiunta da Marco

%            Frecce

\def\arr{\rightarrow }            %  Freccia da applicazione
            %  Freccia da applicazione
   %  Coinplicazione con indent
\def\then{\quad\Rightarrow\quad}      %  Implicazione con indent
\def\QDE{\hbox{\ }\vrule height4pt width4pt depth0pt}                                                              %  Quadratino fine dimostrazione

\def\gBTZ{g_{_{\hbox{\sevenrm BTZ}}}}

\def\np{\vfill\eject}
\def\ni{\noindent}

\def\ss{\vskip 5pt}
\def\ms{\vskip 10pt}
\def\bs{\vskip 15pt}

%%%%%%%%%%%%%%%%%%%%%%%%%%%%%%   BIBLIO.def  %%%%%%%%%%%%%%%%%%%%%%%%%
\bib{\WaldA}{V. Iyer and R. Wald, Phys. Rev. D {\bf 50},  1994, 846}

\bib{\WaldB}{R.M.\ Wald, J.\ Math.\ Phys., {\bf 31}, 2378 (1993) }

\bib{\WaldC}{B.\ S.\ Kay, R.\ M.\ Wald, Physics Reports {\bf 207}, (2),
 49 (1991)}

\bib{\WaldD}{ I.\ Racz, R.\ M.\ Wald, Class. Quantum Grav. {\bf 9}, 2643 (1992)} 

%%%%%%%%%%%%%%%%%%%%%%%%%%%%%%%%%%%%%%%

\bib{\CADM}{M.\ Ferraris and M.\ Francaviglia, Atti Sem. Mat. Univ. Modena, {\bf 37},
61 (1989)} 

\bib{\CADMB}{M.\ Ferraris, M.\ Francaviglia and I.\ Sinicco, Il Nuovo Cimento, {\bf 107B},
(11), 1303 (1992)}

\bib{\CADMC}{M.\ Ferraris and M.\ Francaviglia, Gen.\ Rel.\ Grav., {\bf 22}, (9), 965 (1990)}

\bib{\Lagrange}{M. Ferraris, M. Francaviglia,
in: {\it Mechanics, Analysis and Geometry: 200 Years after Lagrange},
 M. Francaviglia ed., (Elsevier Science Publishers B.V., Amsterdam, 1991)}

\bib{\Cavalese}{M. Ferraris and M. Francaviglia, in: {\it 8th Italian
Conference on General Relativity and Gravitational Physics}, Cavalese (Trento), August 30 --
September 3 (World Scientific, Singapore, 1988)}

\bib{\Robutti}{M.\ Ferraris, M.\ Francaviglia and O.\ Robutti, in:{\it G\'eom\'etrie et Physique},
Proceedings of the {\it Journ\'ees Relativistes 1985} (Marseille, 1985), 112 -- 125; Y.\
Choquet-Bruhat, B.\ Coll, R.\ Kerner, A.\ Lichnerowicz eds. (Hermann, Paris, 1987)}

\bib{\Kolar}{I.\ Kol{\'a}{\v r}, P.\ W.\ Michor, J.\ Slov{\'a}k, 
{\it Natural Operations in Differential Geometry} 
(Springer--Verlag, New York, 1993)}

%%%%%%%%%%%%%%%%%%%%%%%%%%%%%%%%%%%%

\bib{\RevA}{A. Trautman, in: {\it Gravitation: An Introduction to Current Research}, L. Witten ed.
(Wiley, New York, 1962) p. 168}

\bib{\Katz}{J.\ Katz, Class.\ Quantum Grav., {\bf 2}, 423 (1985)}

%%%%%%%%%%%%%%%%%%%%%%%%%%%%%%%%%%%%%%%%%%%%%%%%%%%%%%%%%%%%%%%%%%%%%%
\bib{\BTZRef}{M.\ Ba\~nados, C. Teitelboim, J.\ Zanelli, Phys.\ Rev.\ Lett.\ {\bf 69}, 1849
(1992); M.\ Ba\~nados, M.\ Hennaux, C.\ Teitelboim, J.\ Zanelli, Phys.\ Rev.\ D{\bf 48}
(4), 1506 (1993)}

\bib{\BTZRefB}{J.\ D.\ Brown, J.\ Creighton, R.\ B.\ Mann, Phys. Rev. D{\bf 50}, 6394 (1994)}

\bib{\BTZRefC}{D.\ Cangemi, M.\ Leblanc, R.\ B.\ Mann, Phys. Rev. D{\bf 48}, 3606 (1993)}

\bib{\BTZRefD}{S.\ Carlip, C.\ Teiteilboim, Phys.\ Rev.\ D{\bf 51} (2), 622 (1995)}

\bib{\BTZMann}{S.\ Carlip, J.\ Gegenberg, R.\ B.\ Mann, Phys.\ Rev.\ D{\bf 51} (12), 6854 (1995)}

\bib{\CaGe}{S.\ Carlip, J.\ Gegenberg, Phys.\ Rev.\ D{\bf 2} (44), 424 (1991)}

\bib{\Remarks}{L.\ Fatibene, M.\ Ferraris, M.\ Francaviglia, M.\ Raiteri, Ann.\ Phys.\ (in press);
e-archive: hep-th/9810039}

\bib{\Hawking}{S.\ W.\ Hawking, C.\ J.\ Hunter, D.\ N.\ Page, hep-th/9809035;
S.\ W.\ Hawking, C.\ J.\ Hunter, hep-th/9808085;
C.\ J.\ Hunter, hep-th/9807010;
S.\ W.\ Hawking, C. J. Hunter, Class.\ Quantum Grav.\ {\bf 13}, 2735 (1996);
S.\ W.\ Hawking, G.\ T.\ Hurowitz, Class.\ Quantum Grav.\ {\bf 13}, 1487 (1996);
D.\ N.\ Page, Phys.\ Lett.\ {\bf 78B}, 249 (1978)
}

\bib{\BrownYork}{J.\ D.\ Brown, J.\ W.\ York, Phys.\ Rev.\ D{\bf 47} (4), 1407 (1993)}

\bib{\Brown}{J.\ D.\ Brown, gr-qc/9506085}

\bib{\MannMann}{R.\ B.\ Mann, hep-th/9903229; R.\ B.\ Mann, hep-th/9904148}

\bib{\TAUBNUT}{L.\ Fatibene, M.\ Ferraris, M.\ Francaviglia, M.\ Raiteri, {\it On the Entropy of
Taub-NUT Black Hole Solutions}, (in preparation)}

\bib{\Komar}{A.\ Komar, Phys. Rev. {\bf 113}, 934 (1959)}

%%%%%%%%%%%%%%%%%%%%%%%%%%%%%%%%%%%%%%%%%%%%%%%%%%%%%%%%%%%%%%%%%%%%%%
%%%%%%%%%%%%%%%%%%%%%%%%%%%%%%%%%%%%%%%%%%%%%%%%%%%%%%%%%%%%%%%%%%%%%%

%%%%%%%%%%%%%%%%%%%%%%%%%%%%%%%%%%%%%%%%%%%%%%%%%%%%%%%%%%%%%%%%%%%%%%%%%

\def\titolo{Remarks on Conserved Quantities and Entropy of BTZ Black Hole Solutions. Part I: the
General Setting}
\def\autori{%
L. Fatibene, M. Ferraris, M. Francaviglia, M. Raiteri
}

\def\indirizzo{Dipartimento di Matematica, Universit\`a di Torino,\goodbreak
 via C. Alberto 10, 10123 Torino Italy}

\def\Abstract{
The BTZ stationary black hole solution is considered and its mass and angular momentum are
calculated by means of N\"other theorem. In particular, relative conserved quantities with respect
to a suitably fixed background are discussed. Entropy is then computed in a  geometric  and
macroscopic framework, so that it satisfies the first principle of thermodynamics. In order to
compare this more general framework to the prescription by Wald et al.\ we construct the  maximal  
extension of the BTZ horizon by means of Kruskal--like coordinates.  A discussion about the
different features of the two methods for computing entropy is finally developed.
}

\def\piedipagina{}

\vglue 53.3pt
\raggedbottom

\MainHead

\np
\pagenumbers

%%%%%%%%%%%%%%%%%%%%%%%%%%%%%%%%%%%%%%%%%%%%%%%%%%%%%%%%%%%%%%%%%%%%%%
\NewSection{Introduction}

Since its discovery in 1992, the so-called BTZ black hole solution has often been used in current
literature as a simple but realistic model for black hole physics (see \ref{\BTZRef},
\ref{\BTZRefC}, \ref{\BTZRefB}, \ref{\BTZRefD}, \ref{\BTZMann} and references quoted therein). 
Specifically, it has been assumed as a {\it test model} for both
quantum gravity and for problems related to black hole entropy.
Recently, the BTZ solution has been shown to be the {\it effective solution} in a dimensional
reduced model and many black holes ensuing from string theory are described in terms of the BTZ
solution, at least near the horizon.

Certainly, many facts about entropy remain to be understood from both a statistical and a
geometrical viewpoint. In recent investigations of ours we therefore aimed to review what is known
from a geometrical macroscopic viewpoint and to add some considerations which, as far as we
know, are new in the literature.

The results of our investigations will be presented in two papers, whereby the material has been
divided on the basis of coherence and shortness considerations. We shall refer to them as Part I
(the present paper) and Part II (forthcoming). 
In Part I we consider the standard BTZ solution, seen as a vacuum solution for standard
$(2+1)$-General Relativity with (negative) cosmological constant. We review and apply to the
specific example some methods for defining conserved quantities (as is well known, there are
several methods in literature and it is almost impossible to review all of them in a single paper;
hereafter we shall follow the covariant approach to conserved quantities based on N\"other theorem
- see \ref{\RevA}, \ref{\Lagrange}, \ref{\CADM}, \ref{\CADMB}, \ref{\CADMC}, \ref{\Remarks}).
Moreover a proposal for fixing a background spacetime is suggested in order to correct
so-called {\it anomalous factors} (see \ref{\Katz}) and to produce the expected values of conserved
quantities.

Then we calculate the entropy by relying on a geometrical and global framework
presented in \ref{\Remarks}. As it was already discussed there, our general method contains 
the proposal of Wald et al. as a particular case (see \ref{\WaldA}, \ref{\WaldB}, \ref{\WaldC} and
references quoted therein). Here we apply on purpose also the original Wald's recipe to show that
it allows to obtain the same results but by a much longer route. We believe in fact that a serious
comparison between the two methods is important also because from a theoretical viewpoint Wald's
framework requires in fact additional hypotheses with respect to ours: basically involving the
surface gravity
$\ka$ which in our more general framework is not required to vanish in order to ensure the
horizon to be a {\it bifurcate Killing horizon} (see
\ref{\WaldA}, \ref{\WaldB}). In the BTZ solution these additional requirements of Wald hold
true (with the exception of the extremal case $r_+=r_-$). Thence the entropy calculated by using
Wald's recipe has necessarily to agree with our computation. Other examples in which Wald's
additional hypotheses do not instead hold (such as the Taub-NUT solutions) will be considered
elsewhere (see \ref{\TAUBNUT}), whereby we shall show that our method works also when Wald's
recipe fails for lack of properties of the concerned solution (thus providing a geometrical recipe
for the correct entropy which can be calculated on a statistical basis as in \ref{\Hawking}).

In Part II we shall analyse a triad-affine theory with topological matter
but no cosmological constant, which is called {\it BCEA} and describes the BTZ spacetimes.
This theory has already appeared in the literature (see \ref{\BTZMann}, \ref{\CaGe}).
It has been there shown that it exhibits an ``exchange behaviour'' between conserved quantities
(the total mass, i.e.\ the N\"other conserved quantity associated to a timelike vector is the
parameter $J$ which should correspond to the angular momentum of BTZ spacetimes when described in
standard General Relativity). For what concerns the entropy an   exchange of inner and outer
horizons has also been noticed (see \ref{\BTZMann}). In Part II we shall first obtain the same
results in a geometrical and global formalism. Then we shall explicitely build a  purely metric
(natural) theory, which we shall call {\it BCG theory}, equivalent to the triad-affine BCEA. The
BCG theory can be obtained through a generalized ``dual'' Legendre transformation, which  may be
viewed as a generalization of the Palatini first order variational approach to General Relativity.
Starting from the BCG Lagrangian conserved quantities and entropy will be again calculated for BTZ
spacetime. The results so obtained, in our opinion, will enlight some of the ambiguities present in
the BCEA theory. In particular they allow to truly isolate the ``matter'' contributions to
conserved quantities from the purely gravitational contribution, so to better explain the
calculation previously performed by others in BCEA theory (see \ref{\BTZMann}).

\NewSection{The BTZ solution}

Let us consider a spacetime manifold $M$ (for the moment of arbitrary dimension $n=\dim(M)$) and
the bundle
$\Lor(M)$ of Lorentzian metrics over $M$.
Let us fix a trivialization and denote by $g_{\mu\nu}$ the coefficients of the metric field (as
well used as coordinates on the fibers of $\Lor(M)$), by $\ga^\la_{\mu\nu}$ the Christoffel symbols
(i.e.\ the coefficients of the Levi-Civita connection of the metric $g$), by $r_{\mu\nu}$
the Ricci tensor and by $r$ the scalar curvature (of the metric $g$).

The Hilbert-Einstein Lagrangian with
negative cosmological constant $\La=-1/l^2$ is
$$
L=\calL\>\ds={\al}(r-2\La)\>\sqrt{g}\>\ds
\fl{\LagCCm}$$
where $\ds=\d x^1\land \d x^2 \land \dots \land \d x^n$ is the standard basis for $n$-forms over
$M$ and $\al\not= 0$ is a coupling constant. To compare with results in \ref{\BTZMann} one has to
set
$\al={1\over 2}$. Let us denote the {\it covariant naive momenta} by
$$
\eqalign{
&\pi_{\mu\nu}={\del\calL\over\del g^{\mu\nu}}= \al\sqrt{g}\>(r_{\mu\nu}-\hbox{$1\over 2$}r
g_{\mu\nu} +\La g_{\mu\nu})\cr
&p^{\mu\nu}={\del\calL\over\del r_{\mu\nu}}=\al\sqrt{g} \>g^{\mu\nu}\cr
}
\fn$$
so that for the variation of the Lagrangian $\LagCCm$ we have
$$
\de L=\pi_{\mu\nu}\de g^{\mu\nu} + p^{\mu\nu} \de r_{\mu\nu}
\fl{\FVP}$$

As is well known, $\pi_{\mu\nu}=0$ are the Euler-Lagrange equations for the the Lagrangian
$\LagCCm$, i.e.\ Einstein field equations.

In dimension $3$ there is a $2$-parameter family of black hole solutions (called BTZ
black holes) given by (see \ref{\BTZRef})
$$
\gBTZ=-N^2\dt^2+N^{-2}\dr^2+r^2(N_\phi\dt+d\phi)^2
\fl{\BHMann}$$ 
where we set
$$
\eqalign{
&N^2=-\mu+{r^2\over l^2} + {J^2\over 4r^2} \cr
&N_\phi=-{J\over 2r^2}\cr
}\qquad\qquad\qquad
\eqalign{
&\mu={r^2_++r^2_-\over l^2} \cr
&J=2{r_+r_-\over l}\cr
}
\fn$$
We recall that it has been shown  in \ref{\BTZRefC}  that $\mu$ and $J$ are respectively the ADM
mass and angular momentum at infinity.
One can thence apply various methods to compute the conserved quantities
and the entropy via N\"other's theorem.
In the sequel we shall summarize and compare various approaches.

\NewSection{N\"other Theorem}

The Lagrangian $\LagCCm$ is covariant with respect to the action of diffeomorphisms of spacetime
$M$.  Infinitesimally this is expressed by the following identity which holds for any vector field
$\xi$ over $M$
$$
\d(i_\xi L)= \pi_{\mu\nu}\Lie_\xi g^{\mu\nu}+ p^{\mu\nu}\Lie_\xi r_{\mu\nu}
\fl{\FondID}$$
where $\Lie_\xi$ denotes the Lie derivative operator and $i_\xi$ the contraction (or inner product)
of forms along
$\xi$. By expanding the Lie derivative of the Ricci tensor both field equations
and the N\"other conserved current can be found. In fact, by defining 
$$
u^\la_{\mu\nu}:= \ga^\la_{\mu\nu} - \de^\la_{(\mu} \ga_{\nu)}
\qquad\qquad
 \ga_{\mu}:=\ga^\ep_{\ep\mu}
\fn$$
the Lie derivative of the Ricci tensor can be expressed as follows
$$
\Lie_\xi r_{\mu\nu}=\na_\la \Lie_\xi \big( u^\la_{\mu\nu}\big)
\fn$$ 
Thence we can recast eq.\ $\FondID$ as follows
$$
\Div\>\calE(L,\xi)=\calW(L,\xi)
\fl{\WeakConservation}$$
with
$$
\eqalign{
&\calE(L,\xi)=(p^{\mu\nu}\Lie_\xi u^\la_{\mu\nu}-\calL\xi^\la)\ds_\la\cr
&\calW(L,\xi)=-(\pi_{\mu\nu}\Lie_\xi g^{\mu\nu})\ds\cr
}
\fl{\NotherCorrents}$$
Here $\ds_\mu=i_{\del_\mu}\>\ds$ is the standard basis for $(n-1)$-forms;
$\Div$ denotes the formal divergence operator which acts on forms depending on $k$
derivatives of fields according to the general rule
$$
(j^{k+1}\si)^\ast\Div(\om)=\d((j^k\si)^\ast\om)
\fn$$
where $\d$ is the standard differential of forms over
$M$, $j^k$ denotes derivation up to order $k$ and $\si$ denotes any section of the configuration
bundle (in our case the bundle is $\Lor(M)$ while $k$ is usually $1$ or $2$, depending on how many
derivatives of
$g$ enter $\om$; recall that $g$ enters the Lagrangian and the theory through its second order
derivatives appearing in the curvature tensor). For functions one has $\Div F=(\d_\mu F)\>\d
x^\mu$, where the differential operators
$\d_\mu$ are called {\it total formal derivatives}.

Following the general prescription of \ref{\Lagrange} and \ref{\Robutti},
by computing $\NotherCorrents$ along any configuration $\si$, we can
define the {\it currents}
$\calE(L,\xi,g)$ and $\calW(L,\xi,g)$ on $M$. If $g$ is a solution then
$\calW(L,\xi,g)=0$ and $\calE(L,\xi,g)$ is conserved, i.e.\ it is a closed form on $M$.
Using Bianchi identities in eq.\ $\WeakConservation$ and integrating by parts, we can
({\it algorithmically}) recast (see \ref{\Lagrange}, \ref{\Remarks}, \ref{\Robutti}) the current
$\calE(L,\xi)$ as $$
\eqalign{
&\calE(L,\xi)=\tilde\calE(L,\xi)+\Div\>\calU(L,\xi)\cr
&\tilde\calE(L,\xi)=2\al\sqrt{g}(r_{\mu\nu}-{1\over 2} r g_{\mu\nu}+\La g_{\mu\nu})
g^{\mu\la}\xi^\nu\>\ds_\la\cr
&\calU(L,\xi)=\al\sqrt{g}\> \na^\mu\xi^\nu \ds_{\nu\mu}\cr}
\fl{\Currents}$$
where $\ds_{\nu\mu}=i_{\del_\mu}\>\ds_\nu$ is the standard basis for $(n-2)$-forms over $M$.
Again the current $\tilde\calE(L,\xi,g)=(j^{2k}g)^\ast\tilde\calE(L,\xi)$
vanishes along solutions of field equations, while
$\calE(L,\xi,g)-\tilde\calE(L,\xi,g)=\d\>\calU(L,\xi,g)=(j^{2k}g)^\ast\Div\>\calU(L,\xi)$
(being exact) is {\it strongly conserved}, i.e.\ it is conserved along any configuration $g$ (not
necessarily a solution of field equations).
The current $\calU(L,\xi)$ is known as a {\it superpotential}; for the Lagrangian $\LagCCm$ (with
or without cosmological constant), its value $\Currents$ is also known as Komar potential
(see \ref{\Komar}). Once we specify a covariant Lagrangian $L$ and a vector field $\xi$ on
$M$, a conserved current $\calE(L,\xi)$ and a superpotential $\calU(L,\xi)$ are defined;
moreover, once we specify a configuration $g$, we can compute them on $g$ obtaining
$\calE(L,\xi,g)$ and $\calU(L,\xi,g)$.
Now, let a {\it region} $D$ be a compact submanifold with a boundary $\del D$ which is again a
compact submanifold of $M$; if $g$ is a solution of field equations,
the {\it conserved quantity $Q_D(L,\xi,g)$} is defined as the integral over a region $D$ of
codimension $1$ of the current $\calE(L,\xi,g)$, or equivalently as the integral of
$\calU(L,\xi,g)$ over the boundary $\del D$ of $D$ because of $\Currents$.
In applications one usually chooses $D$ to be a spacelike slice of the ADM
splitting such that $\del D$ is (a branch of) the spatial infinity.
We stress that all quantities defined in $\Currents$ are
$\R$-linear with respect to the vector field $\xi$.
We also stress that $\xi$ is by no means required to be a Killing vector of the metric $g$.

Reverting to the BTZ solution $\BHMann$, notice that the integral over the {\it ``sphere''} $S^1_r$
of radius $r$ of the superpotential
$\Currents$ for the vector fields $\del_t$ and $\del_\phi$ gives
respectively:
$$
\eqalign{
&Q(L,\del_t,\gBTZ)= \int_{S^1_r} \calU(L,\del_t,\gBTZ)=4\pi\al\>{ r^2\over l^2} \cr
&Q(L,\del_\phi,\gBTZ)= \int_{S^1_r} \calU(L,\del_\phi,\gBTZ)=-2\pi\al \> J\cr
}
\fn$$
The {\it mass} $Q(L,\del_t,\gBTZ)$ diverges as one considers the limit $r\rightarrow \infty$;
this is a problem analogous to the well known {\it anomalous factor problem} which Komar potential
is known to be affected of (see \ref{\Katz}).
In other words the integral of the superpotential does not give the correct mass though it gives
the expected {\it angular momentum} $Q(L,\del_\phi,\gBTZ)$.
Here the anomalous factor problem is even worse than for Kerr-Newman metrics or for other
asymptotically flat stationary solutions (see \ref{\CADM}, \ref{\CADMB}, \ref{\CADMC}).
In fact, while there it was just a matter of a wrong factor to be corrected, here it is
primarily a divergence to be cured.
This divergence is typically due to the fact that the BTZ solution is an asymptotically anti-de
Sitter spacetime and the magnitude of the timelike vector field $\del_t$ diverges as it approaches
infinity.

There are (at least) two different possibilities to overcome this situation. First, one may 
define the {\it total conserved quantities} without any reference to their {\it  densities}. 
On the other hand, one can define the conserved quantity {\it in a region $D$} as the integral over
$D$ of its density; then the total conserved quantity is obtained by taking the limit to the
whole spacelike slice, provided that the method converges to a finite result.

\bs
\NewSection{The total conserved quantities}

Let us associate to any vertical vector field $X=\big(\de g^{\mu\nu}\big)\>{\del/\del g^{\mu\nu}}$
over the bundle $\Lor(M)$, i.e.\ for any variation $\de g^{\mu\nu}$ of the (inverse) metric, an
$(n-1)$-form:
$$
\F(L,g)[X]=\al\>(g^{\la\rho}g_{\mu\nu}-\de^\la_{(\mu}\de^\rho_{\nu)})
\na_\rho \big(\de g^{\mu\nu}\big)\>\sqrt{g}\>\ds_\la
\fl{\MPC}$$
where the section $g$ of $\Lor(M)$, i.e.\ a Lorentzian metric, is not required to be a
solution of field equantions. The correspondence $\F(L,g)$ is called {\it Poincar\'e-Cartan
morphism}. Recalling (see \ref{\Kolar} and references quoted therein) that Lie derivatives
$\Lie_\xi g$ of sections $g$ can be interpreted as vertical vectors over $\Lor(M)$
$$
\Lie_\xi g=(\na^\mu\xi^\nu+\na^\nu\xi^\mu){\del\over\del g^{\mu\nu}}
\fn$$
one can rewrite the current $\NotherCorrents$
as:
$$
\calE(L,\xi,g)=\F(L,g)[\Lie_\xi g]-i_\xi L
\fn$$

The form $\F(L,g)$ is irrelevant to field equations,
because it enters a divergence once one integrates by parts equation $\FVP$ and uses $\FondID$.
However, it is tightly related to conserved quantities and N\"other theorem.
These are in fact general features of any field theory (see \ref{\Lagrange}, \ref{\Remarks} and
references quoted therein for the general framework).

The {\it variation of the total conserved quantity} is:
$$
\eqalign{
&\de_XQ_D(L,\xi,g)=\int_D\de_X\calE(L,\xi,g)=\cr
&=\int_D \de_X \big(\F(L,g)[\Lie_\xi g]\big)-\int_D \Lie_\xi\big(\F(L,g)[X]\big)+\cr
&\qquad+\int_{\del D} i_\xi\big(\F(L,g)[X]\big)\cr }
\fn$$
which suggests to us to define the variation of the corrected conserved quantities by
the prescription
$$
\de_X\hat Q_D(L,\xi,g)=\int_{\del D} \Big(\de_X\calU(L,\xi,g)-i_\xi \big(\F(L,g)[X]\big) \Big)
\fl{\VarMass}$$
For the example under investigation, using expressions $\FondID$ and $\MPC$ we get
$$
\de_X\hat Q_D(L,\del_t,\gBTZ) = 2\pi\al\>\de \mu
\qquad
\de_X\hat Q_D(L,\del_\phi,\gBTZ) = -2\pi\al\>\de J
\fn$$
which can be integrated to give the total conserved quantities
$$
\hat Q_D(L,\del_t,\gBTZ)= 2\pi\al\> \mu
\qquad
\hat Q_D(L,\del_\phi,\gBTZ)= -2\pi\al\> J
\fl{\TotalConservedQuantities}$$
We see that this method provides directly the total conserved quantities (as already calculated in
\ref{\BTZRef}) and no extra data are needed other than the solution and the Lagrangian.

In other words, the relevant quantity to replace the Komar potential is
$\de_X\calU(L,\xi,g)-i_\xi\big(\F(L,g)[X]\big)$.
This quantity is uneffected by the addition of divergences to the Lagrangian.
In fact, if one considers a Lagrangian $L'=\Div(\be)$ which is a total divergence one easily
obtains
$\de_X\calU(L',\xi,g)-i_\xi\big(\F(L',g)[X]\big)=0$ identically.

As a second alternative approach, one can instead look for a ($n-2$)-form over $M$ which can be
integrated over the boundary of a region $D$ to give directly the conserved quantity in that
region. This approach relies on the formal integration of equation $\VarMass$.
To perform this task one has to specify some extra information.
First, one fixes some boundary conditions, e.g.\ usually one requires that field variations
vanish on the boundary $\del D$.
Then one seeks for a current $\calB(L,\xi)$ such that, once we set, as usual,
$\calB(L,\xi,g)$ for the pull-back of $\calB(L,\xi)$ along a section $g$, the following holds:
$$
\de_X \calB(L,\xi,g)\Big\vert_{\del D}=i_\xi\big(\F(L,g)[X]\big)\Big\vert_{\del D}
\fl{\ReqA}$$

Usually there is no such a (global and covariant) current $\calB(L,\xi)$.
For example, for standard General Relativity one has
$$
i_\xi\big(\F(L,g)[X]\big)= \de (p^{\mu\nu} u^\la_{\mu\nu}\>\ds_\la)
\fn$$
which does not lead to a possible choice for $\calB(L,\xi)$ because $p^{\mu\nu}
u^\la_{\mu\nu}\>\ds_\la$ is not covariant, i.e.\ it is not a form on the bundle since
$u^\la_{\mu\nu}$ is not a tensor.
To overcome this problem one has to fix some ``coherent'' background connection $\Ga^\la_{\mu\nu}$
(which is assumed to be uneffected by deformations) and define
$$
\calB(L,\xi)=p^{\mu\nu} w^\la_{\mu\nu}\>\xi^\nu\>\ds_{\la\nu}
\qquad\qquad
\left\{
\eqalign{
& w^\la_{\mu\nu}= u^\la_{\mu\nu}- U^\la_{\mu\nu}\cr
& U^\la_{\mu\nu}= \Ga^\la_{\mu\nu}-\de^\la_{(\mu}\Ga_{\nu)}\cr
& \Ga_{\nu}= \Ga^\la_{\nu\la}\cr
}\right.
\fl{\CorrectionADM}$$
In this way the correction term $\calB(L,\xi)$ is covariant; one can recast $\VarMass$ as
$$
\eqalign{
\de_X\hat Q_D(L,\xi,g)=&\int_{\del D} \Big(\de_X\calU(L,\xi,g)-i_\xi\big(\F(L,g)[X]\big)\Big)=
\cr=&\int_{\del D} \de_X[\calU(L,\xi,g)-\calB(L,\xi,g)]\cr
}
\fn$$
that can be formally integrated giving:
$$
\hat Q_D(L,\xi,g)=\int_{\del D} [\calU(L,\xi,g)-\calB(L,\xi,g)]
\fl{\CorrectedConsQuant}$$
We stress that in order to construct a formula like $\CorrectedConsQuant$ a background connection
is needed (or some other ``globalizing'' tool).  The new conserved quantities $\hat Q_D(L,\xi,g)$
depend both on the solution $g_{\mu\nu}$ {\it and} on the backround connection $\Ga^\la_{\mu\nu}$. 
They have thence to be interpreted as the {\it relative} conserved quantities with respect to
$\Ga^\la_{\mu\nu}$.
The physical importance of some background for the theory of conserved quantities was already
recognized in the literature; see for example \ref{\BTZRefB}, \ref{\Hawking}, \ref{\BrownYork}.
As an example, it has been proved (see \ref{\CADMC}, \ref{\Cavalese}) that if one analyses the
(charged) Kerr-Newmann solutions there exist suitable backgrounds which provide {\it reasonable}
mass densities which when integrated on the boundary of a spacelike slice (i.e.\ on spatial
infinity) give the correct total mass.

Of course the choice of $\calB(L,\xi)$ is far not unique.
In particular and for sake of simplicity the background $\Ga^\la_{\mu\nu}$ can be assumed to be the
Levi-Civita connection  of some background metric $h_{\mu\nu}$ (also considered to remain unchanged
under deformations). Then one can add a term which depends just on the background
$(h_{\mu\nu},\Ga^\la_{\mu\nu})$, which, being unchanged under deformations, does not effect
$\ReqA$. To fix such a term one can reasonably require that, if $h_{\mu\nu}$ is also a solution
of fields equations (as it seems physically and mathematically reasonable to require)
the {\it relative} conserved quantities of the background with respect to itself vanish.
This amounts to re-define the correction as follows
$$
\tilde\calB(L,\xi)=[p^{\mu\nu} w^\la_{\mu\nu}\xi^\rho+ \al\>\sqrt{h}\>
h^{\al\rho}{\mathop \na^{(h)}}{}_\al\xi^\la]\>\ds_{\la\rho}
\fl{\NewCorrectionADM}$$
where $\na^{(h)}$ denotes the covariant derivative induced by the background $h$.
Generally speaking, the correction $\NewCorrectionADM$ may be used also in asymptotically flat
solutions when, however, there is a preferred vacuum ($h=\eta=$ Minkowski metric) which, being
flat, reduces $\NewCorrectionADM$ to the simpler correction $\CorrectionADM$.

One can also {\it derive} both the corrected superpotentials $\calU-\calB$ and 
$\calU-\tilde\calB$ as (uncorrected) superpotentials of some suitable Lagrangian. 
In particular $\calU-\calB$ is the superpotential for the {\it first order Lagrangian}
for standard General Relativity (see \ref{\CADMC})
$$
L_1=[\al(r-2\La)\sqrt{g} -d_\la(p^{\mu\nu}w^\la_{\mu\nu})]\>\ds
\fl{\FirstOderLagrangian}$$
where $d_\la$ denotes the formal divergence
while $\calU-\tilde\calB$ is the superpotential for the equivalent Lagrangian
$$
\tilde L_1=[\al(r-2\La)\sqrt{g} -d_\la(p^{\mu\nu}w^\la_{\mu\nu})-\al(R-2\La)\sqrt{h}]\>\ds
\fl{\NewFirstOderLagrangian}$$
where $R$ is the scalar curvature of the background $h$.
In both cases the background has to be considered as a {\it parameter}
so that $\xi$ has to be a Killing vector {\it of the background} (see \ref{\CADMC}).
We remark that both Lagrangians $\FirstOderLagrangian$ and $\NewFirstOderLagrangian$ 
induce a well-defined action functional for a variational principle based on fixing the
metric on the boundary.
We stress that in general these actions differ for surface terms from that used in \ref{\BTZRefB}. 

For the Lagrangian  $\LagCCm$ and the solution $\BHMann$ one can obviously choose as
a background another metric of the same type $\BHMann$ with fixed values $(\mu_0,J_0)$
as parameters.
By a direct computation we find for the corrected superpotential $\calU-\tilde\calB$ 
the conserved quantity
$$
\hat Q(L,\del_t,\gBTZ)= 2\pi\al\>(\mu-\mu_0)
\qquad
\hat Q(L,\del_\phi,\gBTZ)= -2\pi\al\>(J-J_0)
\fn$$
Here the Komar potential of the background in $\NewCorrectionADM$ is essential in order to
{\it cure} the quadratic divergence of the Komar potential of the solution.
This background fixing contains, as particular cases, the backgrounds usually adopted in
literature (see, for example, \ref{\BTZRef}, \ref{\BTZRefC}, \ref{\BTZRefB}).
In particular, the limit ($\mu\arr 0$, $J\arr 0$) corresponds to the {\it vacuum state} in which
the black hole disappears.
Another allowed choice analysed in literature (see \ref{\BTZRefB}) is the anti-de Sitter spacetime
which corresponds to the different limit ($\mu\arr -1$, $J\arr 0$).

\NewSection{Entropy}

The entropy of a (black hole) solution is defined to be a quantity that satisfies
the first principle of thermodynamics:
$$
\de_X \mu = T\de_X\calS+\Om\>\de_X\calJ
\fl{\fpt}$$
for any variation $X$ tangent to the space of solutions, i.e.\ $X$ has to satisfy linearized
field equations.
Here $T$ and $\Om$ are constant quantities with respect to variations $\de_X$ (namely they are
related to the unperturbed solution).
Usually they are assumed to represent the {\it temperature} and {\it angular velocity
of the horizon} of the black hole, so that $\calS$ can be interpreted as the physical entropy of
the system.
The physical value of these parameters has to be provided by physical
arguments, since they are almost undetermined in the present context
(see \ref{\BTZRef}, \ref{\BTZRefD}, \ref{\BTZMann}, \ref{\MannMann}, \ref{\Brown}).
Of course, one can compute one of them out of the others by requiring that
equation $\fpt$ is integrable so that there exists a state function
$\calS$ to fullfil the first principle of thermodynamics.
However, other parameters have to be provided by physical arguments (e.g.\
$T$ has to be the temperature of Hawking radiation).
The ultimate meaning of the work terms in $\fpt$ is that, for example,
$\Om\de\calJ$ is the change in the total mass along an isoentropic transformation.
It has been shown elsewhere (see \ref{\BTZRef}, \ref{\BTZRefD}, \ref{\BTZMann}) that, in order to
make this true,
$\Omega$ has to be the angular velocity of the horizon.
Further terms may in general appear in $\fpt$ due to gauge charges.
Since the example which is here under consideration has no further gauge symmetries we do not
consider these further contributions. 

Of course entropy should also satisfy further requirements (e.g.\ the second principle of
thermodynamics). However, these additional requirements are generally out of control so that
the first principle is what one usually requires with the hope to check the second principle
afterwards. 

By solving $\fpt$ with respect to $\de_X\calS$ and by setting $\xi=\del_t+\Om\>\del_\phi$ one
finds (see \ref{\Remarks})
$$
\eqalign{
\de_X \calS=&{1\over T}[\de_X \mu-\Om\de_X\calJ]=
%\cr=&
{1\over T}\int_\infty\Big(\de_X\calU(L,\xi,g)-i_\xi\big(\F(L,g)[X]\big)\Big)\cr
}
\fl{\fptB}$$
where $\infty$ denotes the spatial infinity of a spacelike slice.
Now one can prove under quite general hypotheses (basically just requiring $\xi$ to be a Killing
vector of the solution $g$) that the integrand quantity
$\de_X\calU(L,\xi,g)-i_\xi\big(\F(L,g)[X]\big)$ is a closed form, so that its
integral does not depend on the integration domain but just on its homology.

Thence one can define the following quantity
$$
\de_X \calS={1\over T}\int_\Si\Big(\de_X\calU(L,\xi,g)-i_\xi\big(\F(L,g)[X]\big)\Big)
\fl{\Entropy}$$
where $\Si$ is {\it any} spatial surface such that $\infty-\Si$ does not enclose
any singularity (in homological notation it is a {\it boundary}).
Then equation $\Entropy$ can be integrated to give a quantity $\calS$ which (because of $\fptB$)
satisfies the first principle of thermodynamics and which is thence a natural candidate to be
interpreted as the entropy.
We remark that we do not need anything but a $1$-parameter family of solutions $g^\ep_{\mu\nu}$
and a Killing vector $\xi$ for the unperturbed solution $g^0_{\mu\nu}$
(here and everywhere we denote by $X$ the infinitesimal generator of the family, which is
a solution of the linearized field equations).
In particular, differently from \ref{\WaldA}, \ref{\WaldB}\ and \ref{\WaldC}, we do not require
anything about maximality of the solution under consideration, anything about horizons and
anything about the vanishing of $\xi$ on horizons (see
\ref{\Remarks}, \ref{\WaldA}). This latter remark is particularly important for the actual
calculations because, as we shall see, it simplifies considerably (both conceptually and
computationally) the expressions involved.

Let now $\ka$ denote the surface gravity so that $T=\ka/(2\pi)$ is the temperature of the Hawking
radiation of BTZ, as shown in \ref{\BTZRefD}, \ref{\MannMann} by means of Euclidean path integrals.
Let  us set $\Om=-N_\phi(r_+)$ which can be shown to be the angular velocity of the BTZ horizon.
We remark that because of the value of $\Om$ the Killing vector $\xi$ becomes null on the horizon
$\calN:(r=r_+)$.
We then easily get
$$
\de_X \calS=8\pi^2\>\al\>\de r_+
\fl{\VarEntropy}$$
which in turn gives
$$
\calS=8\pi^2\>\al\> r_+
\fl{\UnVarEntropy}$$
for the entropy. 

The original Wald's recipe for the entropy needs to choose a particular $\bar \Si$ (the
bifurcation surface) on which the Killing vector
$\xi$, and thence the whole term $i_\xi\big(\F(L,g)[X]\big)$ in $\Entropy$,
vanishes. Of course the existence of such a surface is ensured just in the maximal
extension of the solution.
[For example, for Schwarzschild solution this surface $\bar \Si$ 
corresponds to the surface $U=0$, $V=0$ in Kruskal coordinates.
Notice that the $2$-sphere $U=V=0$ is not covered by spherical coordinates  $(t,r,\te,\phi)$
or outgoing (nor ingoing) coordinates.
To be more precise if one considers any cross section $t=t_0$ and $r=2m$ of the horizon
then $\xi$ does not vanishes on none of it for any value of $t_0$.
Thus Kruskal coordinates are needed in a somehow ``essential'' way to apply Wald's recipe.]

In order to apply the original Wald's recipe to BTZ solution, one should then first build
Kruskal-type coordinates as shown in \ref{\WaldD}.
Once one has verified that $\ka$ is a non-vanishing constant on the horizon
(which is false in the extreme case), this can be done in two steps.
First of all we define Eddington-Finkelstein coordinates
$(u,\rho,\vp)$ which are the parameters along the flows of the vector fields $(\xi,\ze,X)$
we are going to define. The vector field $\xi=\del_t+\Om\del_\phi$ is the Killing vector whose
Killing horizon $\calN$ has to be extended; $u$ is the parameter along its flow; $X$ is a
vector field tangent to $\calN$ such that $X^\la\na_\la u=0$. Finally $\ze$ is a vector field
such that the following conditions have to be satisfied
$$
\left\{
\eqalign{
&\xi^\la \na_\la u=1  \cr
&g(\ze,\xi)=1 \quad \vert\>\ze\>\vert^2=0 \hbox{ on $\calN$}\cr
&g(\ze, X)=0 \cr
}\right.
\fn$$
In these new coordinates the vector fields $(\xi,\ze,X)$ read as
$$
\left\{
\eqalign{
&\xi={\del\over\del u}\cr
&\ze={\del\over\del \rho}\cr
&X={\del\over\del \vp}\cr
}\right.
\fn$$
Notice that even in Eddington-Finkelstein coordinates the vector field $\xi$
does not vanish anywhere in the chart domain.
The BTZ metric reads as
$$
\gBTZ= -f(r(\rho)) \d u^2 + 2 \>\d u\> \d\rho +\Phi(r(\rho)) \>\d u\>\d\vp + \Psi(r(\rho))\d\vp^2
\fn$$
where we set
$$
\left\{
\eqalign{
&f(r)=N^2-r^2(N_\phi+\Om)^2\cr
&\Phi(r)={N^2-r^2(N_\phi^2-\Om^2)\over \Om}\cr
&\Psi(r)=-{N^2-r^2(N_\phi-\Om)^2\over 4\Om^2}\cr
}
\right.
\fn$$
Here $r(\rho)$ is obtained by inverting the change of coordinates, namely
$$
\left\{
\eqalign{
&\rho(r)=\int_{r_+}^{r}{\d {\rm r}\over F({\rm r})}\cr
&F(r)=\sqrt{\Psi(r)\over r^2}\cr
}
\right.
\fn$$
We remark that $F(r)$ is well defined in a neighbourhood of the horizon $\calN$.
Now we can define Kruskal-type coordinates $(U,V,\vp)$ as
$$
\left\{
\eqalign{
&U= e^{\ka u} \cr
&V= -\rho\> e^{-\ka u} \exp\Big(2\ka\int_0^\rho H(\rho)\d\rho\Big)\cr
}\right.
\qquad
H(\rho)={1\over f(r(\rho))}-{1\over 2\ka\rho}
\fl{\FailA}$$
The product $UV$ depends just on $\rho$ and can be regarded as an implicit definition
of $\rho$  as a function $\tilde\rho(UV)$.
Then the BTZ metric reads as
$$
\gBTZ= G(UV) \d U\d V + {\Phi(r(\tilde\rho))\over \ka\> U}\d U\d\vp + \Psi(r(\tilde\rho)) \d\vp^2
\qquad
G(UV)={f(r(\tilde\rho))\over \ka^2\>UV}
\fl{\FailB}$$
In these coordinates the Killing vector $\xi$ reads as
$$
\xi=-\ka\Big(V{\del\over\del V}-U{\del\over\del U}\Big)
\fn$$
which finally vanishes for $\bar\Si:(U=V=0)$.

Thus we have extended the Killing horizon $\calN$ to a bifurcate Killing horizon and we
have identified the bifurcate surface $\bar\Si$.
Now one can ``easily'' compute the entropy:
$$
\de_X S= {1\over T}\int \de_X\>\calU[\xi]_3\d\vp
\quad\mathop{\Longrightarrow}^{\rho=0}_{U=V=0}\quad
\de_X S= 8 \pi^2\>\al\>\de r_+
\then
S= 8 \pi^2 \>\al\>r_+
\fl{\tA}$$
where $\calU[\xi]_3\d\vp$ is the angular part of the superpotential $1$-form $\calU[\xi]$.

We stress that the first expression in $\tA$ for $\de_X S$ is meaningful just on the bifurcate
surface where we can ignore the contribution of the term $i_\xi\big(\F(L,g)[X]\big)$.
We stress moreover that the above method fails in the extreme case.
In fact, in this case $\ka=0$ and the derivation of Kruskal-type coordinates fails at equations
$\FailA$ and $\FailB$ (notice that in this case $f(r)$ identically vanishes).
On the contrary, as we proved (see eq.\ $\VarEntropy$) the second expression in $\tA$ for
$\de_X S$ is correct on {\it any} surface $\Si$. In this way the entropy is not related directly
to a quantity computed on the horizon (see \ref{\Hawking} for a discussion of entropy of Taub-NUT
solutions).

The computations of this Section have been carried out by using MapleV and Tensor package.

\NewSection{Conclusion and Perspectives}

We have determined and discussed the entropy of BTZ solutions.
In our framework the entropy $\fptB$ is clearly related, by its own definition, to N\"other charges
by the first principle of thermodynamics; actually our proposal is to determine {\it a priori}
exactly the quantity that satisfies the first principle of thermodynamics, so that it can be, {\it
a posteriori}, physically interpreted as the entropy.

At first, entropy is a quantity computed at spatial
infinity $\fptB$, as all conserved quantities are. Then one can compute it {\it also} by an
integral on finite regions, provided that the N\"other generator $\xi$ is a Killing vector of the
solution under consideration (see equation $\Entropy$). Finally, if the surface gravity $\ka$ does
not vanish on the horizon, one can extend such an horizon to a bifurcate Killing horizon and
compute integrals on the bifurcate Killing surface on which $\xi$ vanishes (see equation $\tA$).
This latter step is completely useless and computationally boring in applications as well as in the
theoretical framework, as we hope to have shown in the present paper.
Furthermore, our framework, being intrinsecally and geometrically formulated in a global setting,
is in fact valid for a much larger class of theories, namely all field theories with a gauge
invariance, so-called {\it gauge-natural theories} (see \ref{\Remarks}). 
For the same reasons no requirements on signature and/or dimension of spacetime are needed, as
our framework relies only on globality and covariance of the Lagrangian.
We remark that in the original Wald's procedure, $\ka\not=0$ is used for two different purposes,
namely to compute the temperature {\it and} to prove the existence of the bifurcation surfaces.
Non-extremality is essential to the second issue.
Since, in the extreme case, the construction of Kruskal-like
coordinates and bifurcation surface break down (as Wald himself noticed),
we believe that because of this there is little hope to treat the
extreme case through any approaches based on bifurcation surfaces.
We instead believe that our approach, which does not use bifurcation surface, is in a good position 
to treat the extremal cases, too.
Clearly, the extreme cases have to be discussed separately and we hope to address 
the problem in a forthcoming paper.
As we already announced in the Introduction, Part II will revisit the above results in the light
of {\it BCEA theories} (see \ref{\BTZMann} and \ref{\CaGe})
 
Future investigations will be devoted to those cases (e.g.\ Taub-NUT solutions) in which Wald's
prescription for entropy cannot apply at all, as noticed by Wald himself (see \ref{\WaldD}) and
other authors (see also \ref{\Hawking}). In these cases, Wald's prescription fails because of
various reasons, e.g.\ because the orbits of timelike vectors are closed and extra contributions
to the entropy are due to singularities other than those enclosed in Killing horizons (in
particular the {\it Misner string}). Both these reasons prevent the application of the latter
prescription. Since our general prescription does not require the existence of a bifurcate Killing
surface, it allows to overcome these problems, in Taub-NUT solutions .
Results will be published in \ref{\TAUBNUT}, where they will be shown to agree with those found in
\ref{\Hawking} by another formalism.

We finally remark that our formalism is hopefully in a good position to be generalized to
non-stationary black holes, since extra contributions due to non-stationarity seem to be under
control.

\NewSection{Acknowledgments}

We are grateful to I. Volovich for having long ago drawn our attention to the entropy formula of
Wald, as well as to R.\ Mann for having addressed our attention to the BTZ solution and for useful
discussions about it.

\np
\NewSection{References}

\Biblio

\end